\documentclass[11pt]{article}

\usepackage[margin=1in]{geometry}
\usepackage[final]{graphicx}
\usepackage{array}

\usepackage{amsmath,amssymb}
\usepackage{siunitx}
\usepackage{booktabs}
\usepackage{tabularx}
\newcolumntype{L}{>{\raggedright\arraybackslash}X}
\newcolumntype{C}{>{\centering\arraybackslash}X}
\newcolumntype{R}{>{\raggedleft\arraybackslash}X}

\usepackage{float}
\usepackage{changepage}
\usepackage{hyperref}
\usepackage{url}

\hypersetup{
	colorlinks=true,
	linkcolor=blue,
	citecolor=blue,
	urlcolor=blue
}
\newcommand{\authorcontributions}[1]{\section*{Author Contributions}#1}
\newcommand{\funding}[1]{\section*{Funding}#1}
\newcommand{\dataavailability}[1]{\section*{Data Availability}#1}
\newcommand{\acknowledgments}[1]{\section*{Acknowledgments}#1}
\newcommand{\conflictsofinterest}[1]{\section*{Conflicts of Interest}#1}
\newcommand{\abbreviations}[2]{\section*{#1}#2}

\newcommand{\appendixtitles}[1]{}  
\newcommand{\appendixstart}{}      

\newcommand{\PublishersNote}[1]{}  

\newcommand{\Title}[1]{\title{#1}}
\newcommand{\TitleCitation}[1]{}

\newcommand{\orcidauthorA}{} 
\newcommand{\orcidA}{%
	\ifx\orcidauthorA\empty\relax
	\else
	\,\href{https://orcid.org/\orcidauthorA}{\textsuperscript{[ORCID]}}%
	\fi
}

\newcommand{\Author}[1]{\author{#1}}
\newcommand{\AuthorNames}[1]{}
\newcommand{\AuthorCitation}[1]{}

\makeatletter
\newcommand{\@address}{}
\newcommand{\@corres}{}
\newcommand{\address}[1]{\gdef\@address{#1}}
\newcommand{\corres}[1]{\gdef\@corres{#1}}

\newcommand{\printaffiliations}{%
	\begin{center}
		\small
		\@address\par
		\@corres\par
	\end{center}
	\normalsize
}
\makeatother

\newcommand{\keyword}[1]{\noindent\textbf{Keywords:} #1\par}
\makeatletter
\let\orig@abstract\abstract
\let\orig@endabstract\endabstract
\renewcommand{\abstract}[1]{%
	\begin{center}
		\begin{minipage}{0.9\textwidth}
			\orig@abstract
			#1
			\orig@endabstract
		\end{minipage}
	\end{center}
}
\makeatother

\Title{Robust Lifetime Estimation from HPGe Radiation-Sensor Time Series Using Pairwise Ratios and MFV Statistics}

\TitleCitation{Robust Lifetime Estimation from HPGe Radiation-Sensor Time Series Using Pairwise Ratios and MFV Statistics}

\renewcommand{\orcidauthorA}{0000-0003-4605-7937}

\Author{Victor V. Golovko$^{1}$\orcidA{}}

\AuthorNames{Victor V. Golovko}
\AuthorCitation{Golovko, V.V}

\address{$^{1}$ Canadian Nuclear Laboratories, Chalk River, Ontario, Canada}
\corres{Correspondence: victor.golovko@cnl.ca; Tel.: +1-613-584-3311}

\begin{document}
	\maketitle
	\printaffiliations
	
	\abstract{High-purity germanium (HPGe) gamma-ray detectors are core instruments in nuclear physics and astrophysics experiments, where long-term stability and reliable extraction of decay parameters are essential. However, the standard exponential decay analyses of the detector time-series data are often affected by the strong correlations between the fitted parameters and the sensitivity to detector-related fluctuations and  outliers. In this study, we present a robust analysis framework for HPGe detector decay data based on pairwise ratios and the Steiner's most frequent value (MFV) statistic. By forming point-to-point ratios of background-subtracted net counts, the dependence on the absolute detector response is eliminated, removing the amplitude--lifetime correlation inherent to conventional regression. The resulting pairwise lifetime estimates exhibit heavy-tailed behavior, which is efficiently summarized using the MFV, a robust estimator designed for such distributions. For the case study, a long and stable dataset from an HPGe detector was used. This data was gathered during a low-temperature nuclear physics experiment focused on observing the 216 keV gamma-ray line in $^{97}$Ru. Using measurements spanning approximately 10 half-lives, we obtain a mean lifetime of $\tau = 4.0959 \pm 0.0007_{\text{stat}} \pm 0.0110_{\text{syst}}~\text{d}$, corresponding to a half-life of $T_{1/2} = 2.8391 \pm 0.0005_{\text{stat}} \pm 0.0076_{\text{syst}}~\text{d}$. These results demonstrate that the pairwise--MFV approach provides a robust and reproducible tool for analyzing long-duration HPGe detector data in nuclear physics and nuclear astrophysics experiments, particularly for precision decay measurements, detector-stability studies, and low-background monitoring.}
	
	\keyword{HPGe gamma-ray detector; radiation sensors; nuclear physics instrumentation; time-series analysis; robust statistics; exponential decay; uncertainty quantification; Steiner's most frequent value; MFV; non-parametric bootstrap; $^{97}$Ru}

	\section{Introduction}
	
	High-purity germanium (HPGe) gamma-ray detectors are among the most widely used radiation sensors in nuclear physics and astrophysics owing to their excellent energy resolution and suitability for long-duration, low-background measurements. In many applications, including activation studies, decay monitoring, detector-stability campaigns, and precision half-life work, the sensor output is analyzed as a time series that is expected to follow an exponential trend. Therefore, the robust extraction of decay parameters from such detector time series is an instrumentation-relevant problem because backgrounds, outliers, and estimator correlations can bias or destabilize the inferred physical quantities.
	
	Radioactive half-lives are usually measured by recording the activity of a sample over time and fitting an exponential curve. For a nuclide that decays through a single channel, the expected activity, and therefore the expected number of counts in a fixed counting interval, can be written as follows:
	\begin{equation}
		A(t) = A_0\, e^{-t/\tau},
		\label{eq:exp}
	\end{equation}
	where $A_0$ is the activity at $t=0$ and $\tau$ is the mean lifetime. Most analyses estimate $A_0$ and $\tau$ together using non-linear least-squares regression.
	
	Although this approach is widely used, it conceals an important difficulty. The initial activity and the lifetime are often strongly anti-correlated: if the fit tries a slightly longer lifetime, it can compensate by lowering the amplitude so that the curve still passes through the measured points. Consequently, the uncertainty quoted for $\tau$ can depend on how well the absolute scale of the activity is known. Theoretical work by Silverman has shown that this sensitivity does not appear when one works with point-to-point ratios, because the amplitude cancels exactly in that case~\cite{Silverman2014}.
	Lorusso and co-workers later demonstrated that this idea can be successfully applied to real decay data from long-lived radionuclides and that it produces unbiased half-life estimates under realistic experimental conditions~\cite{Lorusso2017}.
	
	Earlier work has also explored ratio-based or non-iterative lifetime extraction approaches from single-exponential decay data~\cite{mangelsdorfConvenientPlotExponential1959}.  For example, difference-equation methods were introduced in the 1970s to eliminate the amplitude analytically and solve directly for the decay constant using small sets of consecutive points~\cite{mooreExperimentalMeasurementExponential1973}.  Related non-iterative schemes were later developed for decay curves with constant background, again relying on algebraic combinations of neighboring data points rather than on global nonlinear regression~\cite{Mukoyama1982}.  In a different context, transform-domain methods based on Laplace projection ratios have been used to estimate lifetimes with improved signal-to-noise properties in single-photon decay spectroscopy~\cite{morenoLifetimeMeasurementUsing1987}.
	Although these approaches share the key idea of suppressing dependence on absolute normalization, they generally treat the resulting lifetime estimates as approximately
	Gaussian and summarize them using conventional least-squares or moment-based arguments.
	This study extends the ratio-based philosophy by explicitly recognizing and addressing the heavy-tailed nature of pairwise lifetime estimators and applying robust statistical tools, such as Steiner's most frequent value (MFV), specifically designed for such distributions.

	From the detector and sensor-analysis perspective, these issues are not restricted to a particular isotope: long-duration radiation-sensor time series frequently exhibit a small number of detector-related anomalies (e.g., local background fluctuations, gain-stability transients, or  peak-fitting outliers) that can overly influence conventional estimators. Therefore, methods that reduce dependence on unknown scale factors and remain stable in the presence of  outliers are directly relevant to instrumentation work in nuclear physics and nuclear astrophysics.
	
	In this study, we follow this complementary idea. Instead of fitting the entire curve at once, we use the information contained in each pair of measured points. For a pure exponential decay, the ratio of the two measurements depends only on the lifetime and not on the initial activity. Therefore, the lifetime can be calculated for every valid pair, and these pairwise lifetimes can be combined into a single estimate.
	
	The distribution of pairwise lifetimes is not Gaussian. Even if the original net counts (or the corresponding count rates) have approximately Gaussian errors at high statistics, the ratio of two noisy quantities produces a heavy-tailed distribution that is close to a Cauchy or Lorentzian shape~\cite{WalckHandbookStatisticalDistributions2007}. These distributions are dominated by long tails and outliers. The arithmetic mean performs very poorly in this situation and even the sample median can have large variance.
	
	Steiner's most frequent value method  was developed for datasets with non-Gaussian and Cauchy-like behaviour~\cite{steinerOptimumMethodsStatistics1997}.
	Instead of minimizing squared residuals, the MFV chooses the location and width of a Cauchy model that minimizes the information lost or Kullback--Leibler (KL) divergence when this model is used to represent the unknown true distribution. In practice, this means that points close to the central cluster carry high weight, whereas distant points are smoothly down-weighted rather than sharply rejected. Therefore, the MFV combines the robustness of the median with a better use of information from the bulk of the data and has been shown, both theoretically and in applications, to give smaller uncertainties than median-based methods~\cite{HajagosSteiner1992,GolovkoMFVBiomolecules}.
	
	The aim of this study is to develop and validate a robust analysis framework for exponential decay time series acquired with high-purity germanium (HPGe) gamma-ray detectors. The framework combines (i) pairwise ratios, which suppress the amplitude--lifetime correlation, and (ii)
	Steiner's most frequent value statistic, which is well suited to summarizing the heavy-tailed distributions arising from ratio-based estimators. The resulting lifetime is compared with a conventional regression analysis and discuss the experimental conditions under which the MFV-based pairwise method is appropriate. 
	
	A dedicated half-life measurement by Goodwin \emph{et al.} has already provided precise values for the electron-capture (EC) half-life of $^{97}$Ru, namely
	$T_{1/2} = 2.8370(14)$~d at room temperature and $T_{1/2} = 2.8382(14)$~d at
	19~K, where the quoted uncertainties represent the total
	(combined) experimental uncertainty, as reported in that study~\cite{GoodwinRu97HL,PhysRevC.80.045501}. In contrast, the present analysis reports statistical and systematic uncertainties separately because the profiling~\cite{PDG2023} and pairwise--MFV procedures allow these contributions to be evaluated independently.
	
	The present study reuses only the 19~K portion of the Goodwin \emph{et al.}~\cite{GoodwinRu97HL,PhysRevC.80.045501} dataset, focusing exclusively on the 216~keV line of $^{97}$Ru. The purpose of this study is not to obtain a more precise half-life but to investigate how different statistical tools behave when applied to a long and stable HPGe detector~\cite{hardyPreciseEfficiencyCalibration2002} time series representative of nuclear-physics instrumentation practice. Methodological studies, such as Golovko’s recent work, have examined how alternative fitting algorithms perform when applied to decay data, providing useful context for understanding estimator dependence~\cite{Golovko_InfScience2025}. In contrast, this analysis concentrates on two specific and well-defined approaches applied to a single, well-characterized dataset. The first is a profile-likelihood analysis~\cite{Rolke2005} of the regression fit, which determines the lifetime uncertainty without relying on the covariance matrix. The second is a pairwise-ratio analysis in which the lifetime is estimated from all point-to-point combinations and summarized using Steiner's most frequent value, a robust estimator designed for unknown distributions. By comparing these two complementary approaches on real sensor data, we aim to clarify their strengths, limitations, and suitability for future precision decay measurements and detector-stability studies in nuclear physics and astrophysics.
	
	This study is intended primarily  as a methodological and instrumentation-focused study. The numerical values obtained for the $^{97}$Ru half-life are secondary and are only used to illustrate the behavior of the estimators.
	
	The remainder of this paper is organized as follows. Section~\ref{sec:ex_data} describes
	$^{97}$Ru 19~K HPGe dataset used in this study, including the measurement conditions, peak-area extraction, and inputs carried forward to the analysis. Section~~\ref{sec:meth} explains the analysis methods in a step-by-step way: we first present conventional regression and residual checks, then introduce the pairwise-ratio lifetime estimator and explain why its distribution is heavy tailed, and finally describe how the MFV and bootstrap procedures are used to produce robust central values and uncertainty estimates. Section~\ref{sec:res} presents the results, compares regression, median-based pairwise estimates, and MFV summaries, and reports the final lifetime and half-life with statistical and systematic components. Section~\ref{sec:dis} discusses why the pairwise--MFV approach is useful for long and stable detector time series, outlines practical conditions under which it performs well, and identifies directions for future work, including potential applications to short-span, high-statistics trigger-rate measurements. Section~\ref{sec:conc} summarizes the main conclusions. Appendix~\ref{app:laplace_plain} provides an independent Laplace-domain consistency check, including the formulas and resampling procedure used to reproduce the cross-check results.

	\section{
		Experimental data
	}
	\label{sec:ex_data}
	
	The measurements analyzed in this study were obtained from the low-temperature dataset acquired at
	19~K during the ruthenium EC study reported by Goodwin \textit{et al.}~\cite{PhysRevC.80.045501,GoodwinRu97HL}.
	In that experiment, the principal published results concerned the room--temperature decay, while the 19~K data were collected in parallel to test for possible temperature dependence.  Only the 19~K dataset is used here.  This study aims to examine how pairwise lifetimes and MFV statistics behave on a single, internally consistent time series in which the experimental conditions remained stable throughout the run.
	
	The sample was a high--purity ruthenium metal disc with a chemical purity better than that of
	\SI{99.999}{\percent}.  It was irradiated with a thermal neutron flux of approximately
	$10^{12}$--$10^{13}$~cm$^{-2}$\,s$^{-1}$ to produce $^{97}$Ru through the $(n,\gamma)$ reaction.  After a short cooling period, the activated disc was mounted directly onto the cryogenic pumping system’s cold head.  In this configuration, the sample reached a stable temperature of $19\pm1$~K and remained in good thermal contact with the cold head for the entire experiment.  The fixed geometry ensured that the sample-detector distance and the detection efficiency did not vary with time.
	
	The activity of $^{97}$Ru was monitored through the 216~keV $\gamma$ ray following
	EC to $^{97}$Tc.  This line is strong, well isolated, and free of significant interference from other activation products.  Spectra were collected with a
	70\%-relative-efficiency coaxial HPGe detector inside a low-background lead shield.  The detector gain and energy calibration were checked regularly and showed no measurable drift, consistent with the detailed stability tests documented in the original experiment~\cite{PhysRevC.80.045501,GoodwinRu97HL}.
	
	The background spectra recorded with the removed sample showed that the continuum around
	216~keV was flat and stable throughout the measurement. Peak fitting was performed using the GF3 routine from the RADWARE suite, which models each peak together with a local polynomial description of the underlying background.  This procedure ensured consistent treatment of small background contributions in every spectrum.
	
	Data were acquired in the fixed-dead-time mode.  The real-time and live time for each spectrum were logged, allowing accurate dead-time corrections.  Counting continued for approximately 30 days, corresponding to roughly ten half-lives of $^{97}$Ru.  For each counting interval the start time, live time, net peak area and its statistical uncertainty were recorded.
	
	In this analysis, we work directly with the background--subtracted net counts reported in the original dataset.  These peak areas represent the total number of events accumulated during each recorded live time of each interval.  In the decay model shown below, the quantity $A(t)$ is interpreted as the expected number of net counts in a fixed counting interval rather than an instantaneous count rate.  Because the live time only slightly varies between spectra, using counts or count rates would lead to identical pairwise ratios up to a constant scaling factor, which is canceled in the pairwise-lifetime formula.
	
	The published analysis in Ref.~\cite{PhysRevC.80.045501,GoodwinRu97HL} concentrated on conventional least--squares regression fits, used primarily to verify detector stability and to compare the 19~K results with the room--temperature run.  No pairwise-ratio or MFV analyses were performed in that work.  Here, we reanalyze the  19~K dataset using pairwise and
	MFV framework without any pre-selection or point removal.  The systematic uncertainty adopted later in this paper arises from our controlled perturbations of the background and timing inputs, not from the systematic budget reported in~\cite{PhysRevC.80.045501,GoodwinRu97HL}. The original study reported no evidence for additional significant systematic in the 19~K data beyond those already accounted for in their regression-based study.
	
\section{Methods
}
\label{sec:meth}

This section explains how we estimate the $^{97}$Ru lifetime from the HPGe net peak areas in a way that is easy to reproduce and robust to rare unstable data points. We start with a standard single-exponential regression to obtain a reference lifetime and verify that the residuals behave like random measurement noise. Then, we apply a pairwise-ratio method that removes the unknown amplitude by construction and converts the time series into a large set of two-point lifetime estimates. Because these pairwise lifetimes form a sharply peaked distribution with long tails, we summarize them with Steiner's most frequent value, which keeps the central cluster and automatically reduces the influence of extreme pairs. Finally, we quantify uncertainty by combining a bootstrap study of statistical variability with a separate, controlled set of perturbations that estimate systematic sensitivity to background, timing, and uncertainty-scale assumptions.

\subsection{Global regression fits
}
\label{regfit}

Before introducing the pairwise method, it is useful to summarize the behavior of standard regression on this dataset.  The decay curve was fitted using two closely related models.  In the first model we treat the initial activity $A_0$ and the lifetime $\tau$ as completely free parameters.  In the second model, we use the physical fact that the activity is proportional to the number of radioactive atoms $N$ through
\begin{equation}
	A(t) = \frac{N}{\tau}\,e^{-t/\tau}.
	\label{eq:N_tau}
\end{equation}
In that parameterization we fit the scale parameter $N$ and the lifetime
$\tau$.

Fits are performed by minimizing the usual chi-squared function
\begin{equation}
	\chi^2 = \sum_k \left(\frac{y_k - A(t_k)}{\sigma_k}\right)^2,
\end{equation}
where $y_k$ and $\sigma_k$ are the measured net peak counts and their statistical uncertainties for the spectrum taken at time $t_k$.
The optimization is carried out in the logarithms of the positive parameters, which improves the numerical stability.  The covariance matrix from the Hessian of $\chi^2$ is used to calculate parameter uncertainties and correlations.

For the standard parameterization the correlation between $\log A_0$
and $\log \tau$ is about $-0.74$, which means that an increase in one parameter is strongly compensated by a decrease in the other.  In the physical parameterization the correlation between $\log N$ and $\log
\tau$ drops to about $+0.24$. Both fits yield essentially the same lifetime, around $\tau \approx 4.095$~d, but the different correlation structure illustrates how sensitive regression can be to the way the model is written.

\subsection{Residual Analysis and Normality Check
}

A regression fit is meaningful only if the residuals behave as expected.
The residual for each point is the difference between the measured net peak counts and the fitted model value for that spectrum.  To check whether these residuals are consistent with the stated statistical uncertainties, we divide each by its individual uncertainty.  If the model is appropriate, these normalized residuals should look like random samples drawn from a standard bell-shaped (normal) distribution.

As a first check, we count how many normalized residuals fall within one and two SDs of zero.  About \SI{68}{\percent} of the points lie inside $\pm1\sigma$ and about \SI{96}{\percent} lie inside
$\pm2\sigma$, which closely matches what is expected for a normal distribution.

To aid visual cross-checking between panels in Fig.~\ref{fig:regression_residuals}, each data point in the upper (regression) plot is rendered with the same color as its residual (\(r_i\)) in the lower panel: points with \(|r_i|\le 1\sigma\) are shown in black, those with \(1\sigma<|r_i|\le 2\sigma\) in blue, and outliers with \(|r_i|>2\sigma\) in red. This one-to-one color mapping makes it immediately clear which time intervals fall within the \(1\sigma\) band, which lie in the \(2\sigma\) band, and which are outside.

To make this assessment more quantitative, we applied three standard tests of normality to the normalized residuals.  Shapiro--Wilk test~\cite{shapiro1965analysis}
gave a $p$-value of $0.64$, the Anderson--Darling test~\cite{anderson1952asymptotic} gave a
$p$-value of $0.54$, and the Lilliefors (Kolmogorov--Smirnov) test~\cite{an1933sulla,smirnov1948table} gave
a $p$-value of $0.52$. Each value is well above the usual
$0.05$ threshold, meaning that none of the tests found evidence that the residuals depart from a normal distribution. Together, these visual and numerical checks confirm that the statistical uncertainties are well-estimated and that the single-exponential model provides an adequate description of the data, consistent with the behavior shown in
Fig.~\ref{fig:regression_residuals}.

\subsection{From the exponential decay to the pairwise lifetimes
}

The pairwise method starts from the same exponential model but uses it differently.  The idea is easiest to see by writing the equation for two different times $t_i$ and $t_j$:
\begin{align}
	A(t_i) &= A_0\, e^{-t_i/\tau},\\
	A(t_j) &= A_0\, e^{-t_j/\tau}.
\end{align}
Taking the ratio of these two expressions removes the unknown $A_0$:
\begin{equation}
	\frac{A(t_i)}{A(t_j)} = e^{-(t_i - t_j)/\tau}.
\end{equation}
Solving this equation for $\tau$ gives
\begin{equation}
	\tau_{ij} = \frac{t_j - t_i}{\ln A(t_i) - \ln A(t_j)}.
	\label{eq:pairwise}
\end{equation}
If the decay really follows a single exponential and the net counts in each interval were measured without error, every pair of times would give the same value of $\tau_{ij}$. In real data, the counts fluctuate statistically, and the detector has a small background. Thus, the calculated pairwise lifetimes form a wide distribution around the true value.

We compute $\tau_{ij}$ for all pairs of points with
$j>i$ and with positive counts.  Pairs with very similar net counts, where $A(t_i) \approx A(t_j)$, lead to massive or even undefined values of $\tau_{ij}$ because the denominator in
Eq.~\eqref{eq:pairwise} becomes small.
These extreme values form the distribution’s long tails.

\subsection{Why is the pairwise distribution Cauchy-like
}
\label{sub:Gauchy}

Even when the original net counts have nearly Gaussian uncertainties
(for example, at high counting statistics), the ratio $A(t_i)/A(t_j)$
does not follow a Gaussian distribution.  When two statistically independent noise variables have standard Gaussian distributions, the distribution of their ratio is exactly of the Cauchy type~\cite{WalckHandbookStatisticalDistributions2007}.
In our case, the measured counts fluctuate approximately Gaussian around non-zero means, and Silverman's exact theory for the two-point estimator show that the resulting probability density develops long tails under realistic high-count conditions~\cite{Silverman2014}.
When Eq.~\eqref{eq:pairwise} is applied to noisy data, the calculated
$\tau_{ij}$ values therefore have a heavy-tailed distribution that is well approximated by a Cauchy form.

For a true Cauchy distribution, neither the mean nor the variance is well defined.  The sample median exists and is often used as a robust summary.  However, the median treats all points in the central half of the data equally and completely ignores how far they lie from the center.  This means that it discards some information that could help tighten the confidence interval.

The MFV method was created with exactly this situation in mind~\cite{steinerOptimumMethodsStatistics1997,GolovkoMFVBiomolecules}.  Instead of relying only on order statistics, the MFV fits a Cauchy-shaped model to the data with unknown distribution in a way that minimizes the loss of information, measured through the Kullback--Leibler divergence between the unknown true distribution and the approximating Cauchy distribution.  In this framework, the central location parameter is identified as the most frequent value, and the scale parameter, sometimes called the dihesion, quantifies the dense core width of the data.

\subsection{Practical MFV algorithm for pairwise lifetimes
}

For a finite sample of pairwise lifetimes $\{\tau_k\}$, the MFV can be computed using an iterative weighted-average procedure. The two key quantities are the location $M$ (the MFV itself) and the scale
$\varepsilon$ (the dihesion). Starting from the initial guesses $M_{0}$
and $\varepsilon_{0}$, where $M_0$ is the sample median and the initial scale chosen~\cite{HajagosSteiner1992} as
\[
\varepsilon_{0} = \frac{\sqrt{3}}{2}\,(\tau_{\max}-\tau_{\min}),
\]
where $\tau_{\max}$ and $\tau_{\min}$ denote the maximum and minimum values of the sample $\{\tau_k\}$, respectively.
The following iteration equations are applied:
\begin{align}
	M_{n+1} &=
	\frac{\displaystyle\sum_k \tau_k \left[\varepsilon_{n}^{2} +
		\left(\tau_k - M_{n}\right)^2\right]^{-1}}
	{\displaystyle\sum_k \left[\varepsilon_{n}^{2} +
		\left(\tau_k - M_{n}\right)^2\right]^{-1}},\\[1ex]
	\varepsilon_{n+1}^{2} &= 3\,
	\frac{\displaystyle\sum_k \left(\tau_k - M_{n}\right)^2
		\left[\varepsilon_{n}^{2} +
		\left(\tau_k - M_{n}\right)^2\right]^{-2}}
	{\displaystyle\sum_k \left[\varepsilon_{n}^{2} +
		\left(\tau_k - M_{n}\right)^2\right]^{-2}}.
\end{align}
These formulas are versions of the continuous expressions derived from the KL minimization with a Cauchy substituting distribution~\cite{steinerOptimumMethodsStatistics1997,GolovkoMFVBiomolecules}.
They can be interpreted as follows: points near the current value of $M$ have large weights, whereas points in the long tails receive negligible weights.  The iterations continue until $M$ and $\varepsilon$ stop changing by more than a small tolerance.  Finally, $M$ is taken as the MFV estimate of the lifetime, and the effective number of central points provides an estimate of its statistical uncertainty.

Once the iterations have converged, the MFV framework also provides a simple internal estimate of the location parameter’s statistical uncertainty.
For each data point, we define a weight
\begin{equation}
	w_k = \frac{\varepsilon^{2}}{\varepsilon^{2} + (\tau_k - M)^{2}},
	\label{eq:mfv_weights}
\end{equation}
which is close to unity for points in the dense central cluster and becomes small in the long tails.  The corresponding effective number of the central points is then
\begin{equation}
	n_{\mathrm{eff}} = \sum_k w_k
	= \sum_k \frac{\varepsilon^{2}}{\varepsilon^{2} + (\tau_k - M)^{2}}.
\end{equation}
Csernyak and Steiner showed that the variance of the MFV estimator can be written in the compact form for a symmetric distribution
\begin{equation}
	\sigma_M = \frac{\varepsilon}{\sqrt{n_{\mathrm{eff}}}},
	\label{eq:mfv_sigma_internal}
\end{equation}
which plays the role of a $1\sigma$ standard uncertainty on $M$~\cite{steinerOptimumMethodsStatistics1997}.
In practical terms, $\varepsilon$ sets the intrinsic width of the central cluster, while $n_{\mathrm{eff}}$ counts how many points carry substantial weight in determining the MFV.  This internal MFV uncertainty is used in this work as the quoted statistical error of the MFV lifetime
(see Sec.~\ref{sec:results_pairwise}), while bootstrap-based intervals are employed as an external consistency check on the heavy-tailed behaviour of the estimator.

As a validation of MFV implementation, we applied the same MFV and bootstrap procedure to the neutron--lifetime measurements used in Ref.~\cite{zhang2022mfv}. We reproduced their published MFV estimate $\tau_{n} = 881.16^{+2.25}_{-2.35}\,\text{s}$ and the corresponding
68.27\% and 95.45\% confidence intervals. We have confirmed the numerical correctness of our MFV and resampling routines.

Hajagos and Steiner compared MFV-based filters with median filters and found that MFV filtering suppresses outliers at least as well as the median while producing smaller random errors in the remaining signal~\cite{HajagosSteiner1992}.  In other words, the MFV maintains the same robustness as the median but achieves better precision because it uses smooth  weights~(see Eq.~\ref{eq:mfv_weights}) instead of a hard cut around the median.

\subsection{Uncertainty estimation for MFV and median
}

To quantify the statistical spread of the robust estimators we use a non-parametric bootstrap applied to the original time series rather than to the $\tau_{ij}$ values themselves. Each bootstrap sample is constructed by resampling the spectra list
$\{t_k, y_k, \sigma_k\}$ with replacement. The resampled spectra are then sorted back into chronological order, all valid pairwise lifetimes $\tau_{ij}$ are recomputed, and both the sample median and the MFV are evaluated for that bootstrap realization.

The ensemble of bootstrap medians is summarized using its central 68.27\%
percentile interval, defined by the 15.865\% and 84.135\% quantiles. This interval would correspond with the usual $\pm1\sigma$ range. We treat it as the percentile-based analogue of a one-standard-deviation confidence interval. The median quoted in the results section is the median of the original (non-resampled) pairwise distribution, while the bootstrap percentiles provide  ``$\pm1\sigma$'' error bars around this central value.

For the MFV we adopt the internal uncertainty returned by the MFV algorithm itself as the primary statistical error. This internal error, denoted $\sigma_M$,
is derived from the effective number of points in the dense central cluster. It provides an efficient variance estimate for datasets with unknown distribution (see Eq.~\ref{eq:mfv_sigma_internal}). The bootstrap distribution of MFV values is used as a diagnostic: its standard deviation and central 68.27\% interval are reported for completeness. However, they are not used as the quoted MFV error in the final lifetime and half-life values.

Although the pairwise-lifetime distribution exhibits long, asymmetric tails, this does not conflict with the assumptions behind the MFV uncertainty estimate.
What matters for the MFV is not the distribution’s global shape but the symmetry of the central region where the MFV weights are large.
Direct quantile checks centered on the MFV value confirm that this core is nearly symmetric.  The 25\,\% and 75\,\% quantiles lie only $0.0501$~d below and
$0.0518$~d above the MFV location ($M=4.0959$~d), a ratio of $1.03$, indicating excellent symmetry in the middle 50\% of the data.
A wider interval, the 68.27\,\% (normal-equivalent $\pm1\sigma$) region, shows slightly larger asymmetry, $0.1106$~d below and $0.1245$~d above $M$
(a ratio of $1.13$) because this range already extends into the long tails.  These tails contain very little MFV weight and therefore have almost no influence on the MFV location or internal uncertainty.

This behavior is directly related to the MFV weights
(Eq.~\ref{eq:mfv_weights}), which assign high weight to points close to $M$ and suppress points farther away.  As a result, both the MFV location and its internal standard uncertainty (Eq.~\ref{eq:mfv_sigma_internal}) are determined almost entirely by the symmetric central cluster, whereas the contributions from the highly asymmetric tails are negligible. For this reason, no trimming or data selection is applied: all pairwise lifetimes enter the MFV calculation, but their influence is regulated by the MFV weighting scheme.

Generally, using the internal MFV uncertainty ($\sigma_M$) assumes that the distribution is approximately symmetric in the central part, which significantly contributes to the MFV weight. If simple symmetry checks, such as comparing the distances from ($M$) to the 25\% and 75\% quantiles or to the bounds of a central 68.27\% interval, show a clearly asymmetric core, it is safer not to rely on Eq.~\eqref{eq:mfv_sigma_internal} as the main error estimate. In these situations, while keeping the MFV as the central location estimator, it is better to derive its statistical uncertainty directly from the bootstrap distribution of MFV values. A sensible approach is using the central 68.27\% bootstrap interval around ($M$) is used as an asymmetric ``$1\sigma$-equivalent'' confidence interval. If a symmetric error bar is desired, use the larger of the upper and lower deviations from that interval. This bootstrap-based method maintains the MFV estimator’s robustness while avoiding dependence on a symmetry assumption that might not be appropriate for a specific dataset. A practical implementation of this approach, even for datasets with fewer than 10 elements, that also takes individual elements' uncertainty into account is shown in Ref.~\cite{Golovko2025sensors}.

\subsection{Systematic uncertainties
}

Systematic effects were evaluated by repeating the standard regression fit under controlled modifications of the input data.
The purpose of this study was to assess how sensitive the fitted lifetime is to reasonable variations in the background level, the time stamps of the measurements, and the estimated statistical uncertainties.
Each of these quantities can, in principle, influence the extracted decay constant. Therefore, each was perturbed by a conservative amount, after which the full regression analysis was repeated.

In the likelihood profiling \cite{Rolke2005} used to determine statistical uncertainty, the background subtracted peak areas remain fixed. The resulting $\Delta\chi^{2}(\tau)$ curve therefore reflects only the statistical scatter of the measured counts.
Any possible background subtraction imperfections must be examined separately.
We shifted all net peak areas upward and downward by a constant amount representing the uncertainty of the background determination. Because the 216~keV continuum is small, flat, and locally fitted for each spectrum, its uncertainty is most naturally characterized by the typical statistical uncertainty of the peak areas themselves. For this reason, the background variation was set equal to the median peak-area uncertainty, approximately $\pm 4.7\times10^{2}$ counts.
Repeating the regression with this modified dataset yields a change in the fitted lifetime of about $0.0110$~d, which is considered the background-related systematic uncertainty.

Two additional sources were examined to account for possible instrumental effects.
A global shift of all time stamps by one minute in either direction was used to represent a conservative estimate of any residual synchronization error in the data acquisition clock. Similarly, all statistical error bars were scaled by 1\% to test the fit's sensitivity to modest imperfections in the uncertainty model, such as those arising from peak-shape assumptions or dead-time corrections. Both of these variations produced changes in the fitted lifetime that were orders of magnitude smaller than the background effect and can be regarded as negligible at the present level of precision.

When combined in quadrature, the systematic contributions give a total systematic uncertainty of about $0.0110$~d on the lifetime (and $0.0076$~d on the half-life). Because the pairwise lifetimes are computed from the same background-corrected peak areas used in the regression fit, the same systematic budget is applied to the MFV result.

The core results in this study are obtained in the time domain using regression, pairwise ratios, MFV summarization, and bootstrap resampling.  To provide an independent cross-check that does not use the same fitting structure, we also repeat the lifetime extraction in the Laplace~\cite{Istratov.Vyvenko1999} domain.  These Laplace-domain checks follow the same guiding idea as the pairwise method: they are built from discrete sums evaluated at the actual sampling times, and one of the two checks uses ratios to cancel out the overall amplitude cancels. Appendix~\ref{app:laplace_plain} provides the explicit formulas, weighting used in the Laplace-space fit, choice of the $s$ grid, and bootstrap procedure used to compute the quoted 68.27\% intervals.

\section{Results
}
\label{sec:res}
The results are organized in three steps.  First, we summarize the behavior of standard regression fits and associated residual checks.  Second, we describe the distribution of all pairwise lifetimes and compare their bootstrap behavior with those of the median and MFV estimators.  Finally, we quote the recommended mean lifetime and half-life obtained by combining the MFV central value, its MFV-based statistical uncertainty, and the regression study’s systematic uncertainty budget.

\begin{figure}[t]
	\centering
	\includegraphics[width=0.95\textwidth]{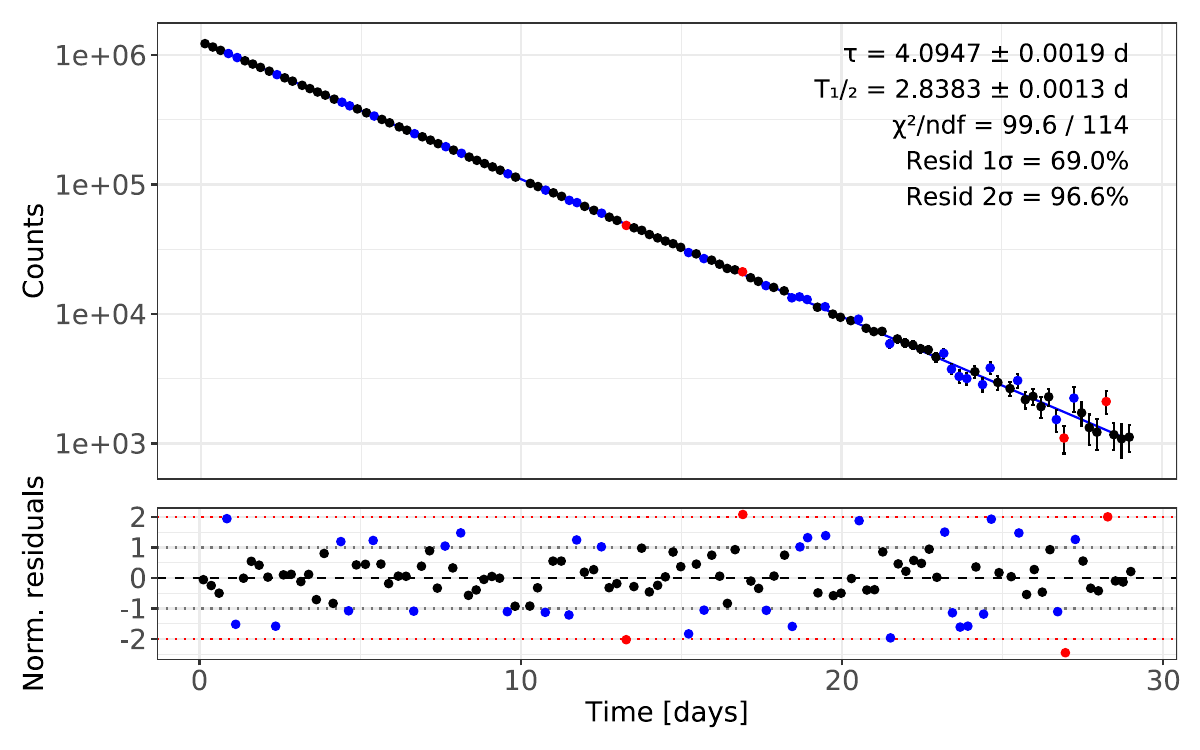}
	\caption{
		Regression analysis of the $^{97}$Ru 19~K dataset.
		The upper panel shows the exponential decay fit (solid line)
		together with the background-corrected net peak counts. The lower panel shows the normalized residuals with
		$\pm1\sigma$ and $\pm2\sigma$ reference bands.
		The absence of structure in the residuals confirms that the
		single-exponential model is appropriate and that the statistical uncertainties are correctly estimated.
	}
	\label{fig:regression_residuals}
\end{figure}

\subsection{Regression Fits, Profiles, and Regressions
}

The standard two-parameter regression of the background-subtracted peak areas yields a best-fit lifetime of the following:
\[
\tau_{\text{reg}} = 4.0947~\text{d},
\]
which corresponds to the half-life of
\[
T_{1/2,\text{reg}} = \tau_{\text{reg}}\ln 2
= 2.8383~\text{d}.
\]
Fits are performed in the logarithms of the positive parameters using the following:
a $\chi^{2}$ minimization as described in Sec.~\ref{regfit}, and the resulting minimum is well behaved.
The fitted curve together with the normalized residuals is shown in Table~\ref{tab:Ru97_uncertainties}.
Fig.~\ref{fig:regression_residuals} illustrates the visual quality of the fit and the absence of time-dependent structure in the residuals.
The overall goodness of fit is characterized by
$\chi^{2}/\text{ndf} = 99.6/114 \approx 0.87$, indicating that a single exponential provides an adequate description of the data within the quoted statistical uncertainties.

To study the dependence on parameterization, we performed two versions of the fit: a ``standard'' form in which $(A_{0},\tau)$ are free parameters (Eq.~\ref{eq:exp}), and a ``physical'' form in which the parameters are $(N,\tau)$ are free parameters (Eq.~\ref{eq:N_tau}).  In the standard parameterization the correlation coefficient between $\log A_{0}$ and $\log\tau$ is about
$-0.74$, indicating a strong anti-correlation between amplitude and lifetime.  In the physical parameterization the correlation between
$\log N$ and $\log\tau$ is reduced in magnitude and changes sign to approximately
$+0.24$. Despite these differences in correlation structure, both parameterizations give virtually identical best-fit lifetimes and very similar curvature of the $\chi^{2}$ surface in the direction of $\tau$.

\begin{figure}[t]
	\centering
	\includegraphics[width=0.85\textwidth]{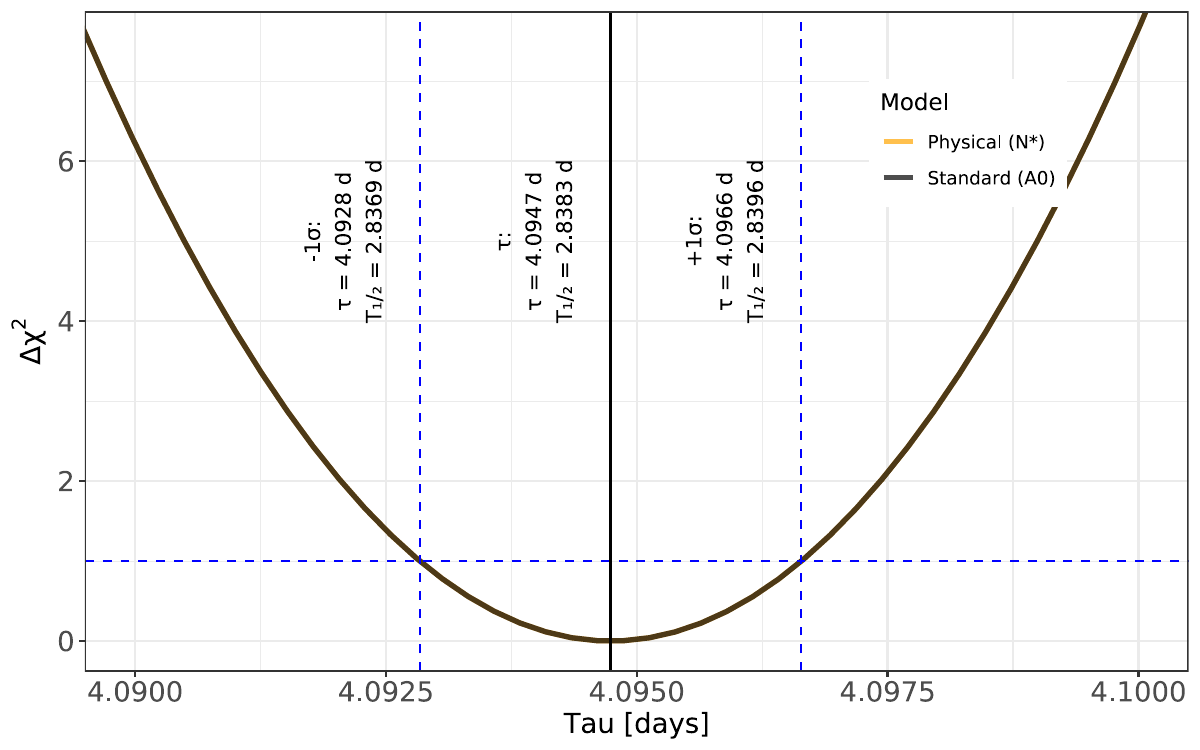}
	\caption{
		Comparison of likelihood profiles for the standard $(A_0,\tau)$ and
		physical $(N,\tau)$ parameterisations.  The $\Delta\chi^{2}(\tau)$ curves nearly overlap, indicating that the extracted lifetime and
		statistical uncertainty are insensitive to the amplitude
		parameter.  The vertical lines mark the best-fit values and the $\pm1\sigma$ range corresponding to $\Delta\chi^{2}=1$.}
	\label{fig:profile_comparison}
\end{figure}

The statistical uncertainty for the regression lifetime is taken from the one-parameter profile likelihood in $\tau$.  For each fixed value of $\tau$ the amplitude parameter is re-optimized and the resulting $\Delta\chi^{2}(\tau) = \chi^{2}(\tau) - \chi^{2}_{\min}$ is constructed.  The half-width of this profile at $\Delta\chi^{2}=1$ defines the $1\sigma$ confidence interval for a single parameter of interest.  Applying this procedure to the standard fit gives the following:
\[
\tau_{\text{reg}} = 4.0947 \pm 0.0019_{\text{stat}}~\text{d},
\]
or equivalently
\[
T_{1/2,\text{reg}} = 2.8383 \pm 0.0013_{\text{stat}}~\text{d}.
\]
The covariance matrix (Hessian) error estimates align with these profile values within a margin of less than one percent. However, we adopt the profile-based uncertainty as our primary regression result because it does not rely on a approximation to the likelihood surface.  Comparison of
$\Delta\chi^{2}(\tau)$ for the standard and physical parameterizations is shown in Fig.~\ref{fig:profile_comparison}. The two curves essentially overlap, confirming that the extracted statistical uncertainty on $\tau$ is insensitive to the amplitude parameter.

Systematic effects associated with the subtracted background, a possible global time offset, and a global rescaling of the statistical error bars are evaluated by repeating the regression fits with perturbed inputs. We shift the background-level up and down by an amount equal to the median statistical uncertainty of the net peak counts ($\pm 4.65\times10^{2}$ counts), and we shift the time axis globally by $\pm 1$~min. We also rescale all quoted uncertainties by $\pm 1$\,\%. For each perturbation, the fit is repeated, and the change in the best-fit lifetime is converted into a systematic uncertainty.  The dominant contribution originates from the background variation, whereas the timing and error-scale effects are negligible at this level of precision.  The resulting uncertainty budget for both the lifetime and half-life is summarized in Table~\ref{tab:Ru97_uncertainties}.

\begin{table}[t]
	\centering
	\small
	\setlength{\tabcolsep}{5pt}
	\caption{
		Uncertainties for the $^{97}$Ru mean lifetime $\tau$ and half-life
		$T_{1/2}$ (days). The relative values are computed with respect to the profile-based central value $\tau = 4.0947$~d.  For each source, nominal value and its $1\sigma$ uncertainty used for the
		$\pm1\sigma$ shifts are shown in the first column.  ``Statistical''
		where denotes the one-parameter profile half-width at $\Delta\chi^{2}=1$.
		The dominant systematic contribution arises from the background-level (BG) variation. The timing and error-scale effects are negligible. The BG uncertainty corresponds to a uniform $\pm 4.65\times10^{2}$-counts shift
		applied to all the net peak areas in the analysis.
	}
	\label{tab:Ru97_uncertainties}
	\begin{tabularx}{\linewidth}{@{} L c c c @{}}
		\toprule
		Component (nominal $\pm$ uncertainty) &
		$\tau$ [days] &
		$T_{1/2}$ [days] &
		Relative (\%) \\
		\midrule
		
		Central value (from profile minimum) &
		4.0947 &
		2.8383 &
		--- \\
		
		Statistical (profile $\Delta \chi^{2} = 1$;
		Best $=4.0947$;
		$1\sigma=[4.0928,\,4.0966]$) &
		$\pm\,0.0019$ &
		$\pm\,0.0013$ &
		$\pm\,0.046$ \\
		
		Systematic: BG level
		[subtracted BG, $0 \pm 4.65\times 10^{2}$ counts] &
		$\pm\,0.0110$ &
		$\pm\,0.0076$ &
		$\pm\,0.269$ \\
		
		Systematic: time offset
		[$0 \pm 1$ min] &
		$\pm\,2.8\times10^{-7}$ &
		$\pm\,1.9\times10^{-7}$ &
		$\approx 0.000007$ \\
		
		Systematic: error-scale factor
		[$1.00 \pm 0.01$] &
		$\approx 0$ &
		$\approx 0$ &
		$\approx 0$ \\
		
		\midrule
		Total systematic (quadrature of systematics) &
		$\pm\,0.0110$ &
		$\pm\,0.0076$ &
		$\pm\,0.269$ \\
		
		Total (stat $\oplus$ systematic) &
		$\pm\,0.0112$ &
		$\pm\,0.0078$ &
		$\pm\,0.273$ \\
		\bottomrule
	\end{tabularx}
\end{table}

The regression fit quality is further assessed through the distribution of normalized residuals, defined as the differences between the measured net peak counts and the fitted model values divided by the individual statistical uncertainties.  Approximately \SI{69}{\percent} of the points lie within $\pm1\sigma$ and about \SI{97}{\percent} lie within $\pm2\sigma$, in excellent agreement with the expectations for a standard normal distribution.  Three classical normality tests applied to the residuals yield $p$-values well above the usual 0.05 threshold: the Shapiro--Wilk test~\cite{shapiro1965analysis} gives $p=0.64$, the Anderson--Darling test~\cite{anderson1952asymptotic} gives
$p=0.54$, and the Lilliefors (Kolmogorov--Smirnov) test~\cite{an1933sulla,smirnov1948table} gives $p=0.52$. With the visual inspection in Fig.~\ref{fig:regression_residuals}, these results confirm that the single-exponential model adequately describes the data and that the quoted statistical uncertainties are neither underestimated nor overestimated.

\subsection{Distribution of the pairwise lifetimes
}

Using Eq.~\eqref{eq:pairwise}, we compute $\tau_{ij}$ for all valid points pairs.  The resulting distribution contains several thousand values and has a pronounced sharp peak near $4.096$~d with long tails extending to both smaller and larger lifetimes.  When plotted as a histogram on a suitable range, the peak is narrow and symmetric.  This shape is typical of a Cauchy-like distribution and matches the expectation (Sec.~\ref{sub:Gauchy}).

Simple sample median of this distribution is
\begin{equation}
	\tau_{\text{med}} = 4.0962~\text{d}.
\end{equation}
Applying the MFV algorithm to the same data gives the following:
\begin{equation}
	\tau_{\text{MFV}} = 4.0959~\text{d}.
\end{equation}
The close agreement between the median and MFV central values indicates that the bulk of the data is well behaved, while the estimated uncertainties are mainly influenced by the long tails.

\begin{figure}[t]
	\centering
	\includegraphics[width=0.95\textwidth]{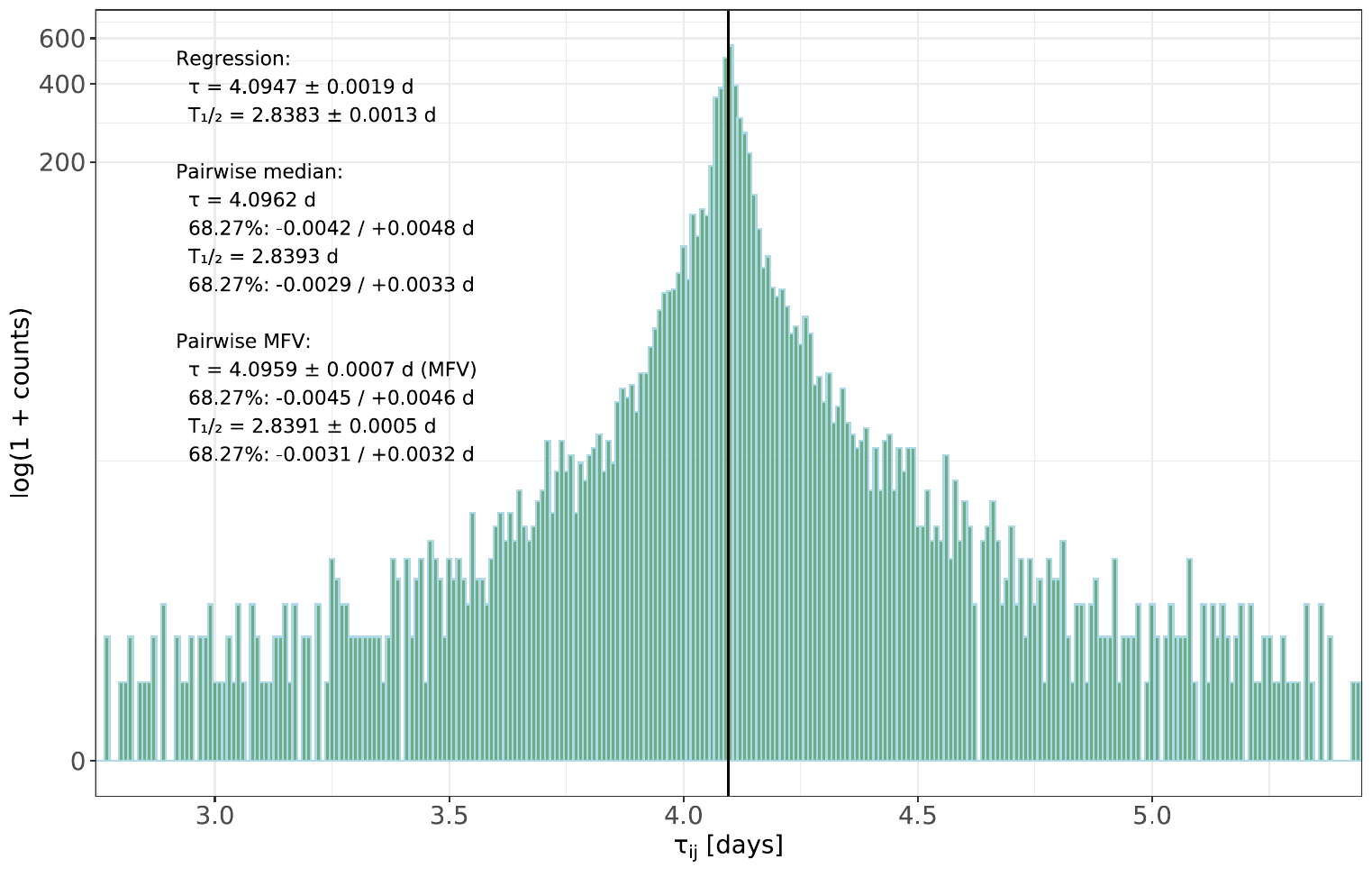}
	\caption{
		Distribution of all pairwise lifetimes $\tau_{ij}$ constructed from the background-corrected net peak counts. Sharp central peak with long heavy tails is characteristic of Cauchy-like behavior. Solid vertical
		line marks the MFV lifetime estimate, while the inset text
		summarizes the regression result, the median and its 68.27\% bootstrap interval, and the MFV value with its MFV-based uncertainty.}
	\label{fig:pairwise_hist}
\end{figure}

The overall shape of the pairwise-lifetime distribution and the relationship between the different estimators are shown in  Fig.~\ref{fig:pairwise_hist}.  The histogram highlights the narrow central peak around $4.096$~d together with the extended wings produced by extreme pairs.  The vertical line marks the regression result, the pairwise distribution median, and the MFV estimate, indicating that all three central values are consistent within their quoted uncertainties.

A bootstrap study with 3,000 resamples, constructed by resampling the original time series, shows that the central 68.27\% of the median distribution
$\tau_{\text{med}} - 0.0042~\text{d}$ to $\tau_{\text{med}} + 0.0048~\text{d}$.
The corresponding half-life interval is given as
$T_{1/2,\text{med}} - 0.0029~\text{d}$ to $T_{1/2,\text{med}} + 0.0033~\text{d}$.
For the MFV estimator the internal MFV uncertainty (Eq.~\ref{eq:mfv_sigma_internal}) is
$\sigma_M \approx 0.0007$~d, corresponding to about $0.0005$~d on the half-life.
The bootstrap MFV values form a wider 68.27\% interval of
$\tau_{\text{MFV}} - 0.0045~\text{d}$ to $\tau_{\text{MFV}} + 0.0046~\text{d}$,
reflecting the influence of extreme pairs on the distribution's long tails, but the central MFV value remains stable. This behavior is in line with the general result that MFV-based estimators efficiently use the central cluster structure while retaining strong resistance to outliers~\cite{HajagosSteiner1992,GolovkoMFVBiomolecules}. We adopt the MFV internal uncertainty as the quoted statistical error because it reflects the information content of the dense central cluster, which dominates the estimator, whereas the bootstrap interval is intentionally sensitive to all data.

\subsection{Final lifetime and half-life
}
\label{sec:results_pairwise}

Combining the MFV central value with statistical uncertainty (Eq.~\ref{eq:mfv_sigma_internal}) and the systematic uncertainty budget from the regression study, we quote the following:
\begin{equation}
	\tau(^{97}\text{Ru}) = 4.0959 \pm 0.0007_{\text{stat}} \pm
	0.0110_{\text{syst}}~\text{d}.
\end{equation}
The corresponding half-life is given by
\begin{equation}
	T_{1/2} = \tau \ln 2 = 2.8391 \pm 0.0005_{\text{stat}} \pm 0.0076_{\text{syst}}~\text{d}.
\end{equation}
These values are fully consistent with the earlier regression-based determinations on the same dataset and with the adopted literature values, but they are obtained with a method that is almost completely insensitive to the amplitude--lifetime correlation and that treats the heavy-tailed nature of the pairwise distribution in a statistically sound way.

As an additional validation of our MFV implementation, we compared our result with an independent MFV--based determination of the \({}^{97}\mathrm{Ru}\) half-life reported recently in Ref.~\cite{Golovko_InfScience2025}. In that study, the MFV and hybrid parametric bootstrapping (HPB) framework was applied not to raw decay curves, but to a compiled set of historical half-life measurements, each treated as an individual value with its published uncertainty. Therefore, the MFV algorithm operated on a small  dataset consisting of previously reported results rather than on a single, internally consistent time series. The resulting half-life was
\begin{equation}
	T_{1/2}^{\mathrm{MFV(HPB)}} = 2.8385^{+0.0022}_{-0.0075}\,\mathrm{d},
\end{equation}
which reflects the statistical consensus of those legacy measurements after uncertainty weighting and MFV-based outlier resistance were applied.

Although this value is not directly comparable to the single-experiment 19~K dataset analyzed in this work, our MFV result lies within its quoted uncertainty band. This agreement serves as a methodological cross-check: it confirms that our MFV procedure reproduces published MFV-based evaluations when supplied with similar types of input data, thereby validating the numerical implementation before applying it to the full pairwise--MFV analysis of the high-precision 19~K decay series.

\section{Discussion and future work
}
\label{sec:dis}

This section summarizes the conceptual foundations of the pairwise MFV method and explains why it provides a robust lifetime estimate for datasets that are well described by a single exponential but exhibit heavy-tailed statistical behaviour.  The discussion is organized around three key ideas: the use of point-to-point ratios, the heavy-tailed nature of the resulting lifetime distribution, and the role of resampling in quantifying uncertainties.
We then outline the practical conditions under which the MFV method performs well and comment on how the resulting uncertainties relate to earlier analyses of the same dataset.

The first principle behind the MFV approach is that ratios of consecutive or widely separated measurements eliminate the unknown initial activity.
Assuming that the decay follows a single exponential over the measurement interval and that detector efficiency, dead-time corrections, and background subtraction remain stable in time, each ratio depends only on the lifetime.
Under these conditions, the method becomes largely insensitive to uncertainties in absolute efficiency or source strength.
This invariance to multiplicative factors ensures that the subsequent analysis focuses directly on the parameter of interest.

The second idea concerns the statistical nature of pairwise lifetimes.
The distribution of $\tau_{ij}$ is known to be heavy-tailed.  In such cases, the arithmetic mean performs poorly because  extreme values dominate it.  The sample median offers much better stability, but it only uses ordering information and ignores the distances between points.
The MFV estimator improves on this by finding the central region of the data that minimizes information loss (KL divergence) between the true, unknown distribution and the Cauchy model~\cite{steinerOptimumMethodsStatistics1997,GolovkoMFVBiomolecules}.
Values near the center contribute strongly, whereas those in the tails are  down-weighted. This produces an estimator that is nearly as robust as the median but with significantly lower variance  datasets~\cite{HajagosSteiner1992}.

The third component of the method uses resampling to evaluate the MFV estimator’s uncertainty.  The bootstrap approach requires no assumptions about the shape of the estimator distribution and performs reliably for skewed or long-tailed data~\cite{efron1994introduction}.
When combined with the MFV, the bootstrap produces confidence intervals that capture both the influence of the heavy tails and the dataset’s finite size.
A natural extension is a hybrid parametric bootstrap in which each resample includes new point selections and new simulated peak areas based on their individual uncertainties.
Such hybrid schemes have already been applied in related nuclear and molecular studies~\cite{Golovko_InfScience2025,GolovkoMFVBiomolecules}.
They may become increasingly valuable for future high-precision lifetime measurements.

This study presents several practical guidelines for applying the MFV method.  The approach performs best when the underlying decay is well described by a single exponential, the experimental setup is stable in time, and a central value that is not unduly influenced by a small number of early, high-statistics points.
If the data show indications of multicomponent decay, time-dependent systematic, or strong correlations between measurement times, then an explicit multicomponent regression model may provide a more accurate description.

A final point concerns the experimental conditions under which the pairwise method is expected to work reliably.  Silverman's analytical treatment shows that the distribution of two-point lifetime estimates approaches a
Cauchy-like form only when the measurement sequence satisfies several practical requirements.  The most important of these is that each spectrum must contain enough counts for the statistical fluctuations in the net peak areas to be well approximated by Gaussian noise.  When the counting statistics are high, the ratio of two measurements behaves in a predictable manner, and the resulting pairwise lifetimes form a sharply peaked distribution rather than a broad, ill-defined cloud of values.

The second requirement is that the dataset must be sufficiently long to provide many independent pairs.  Silverman's calculations indicate that when the time series contains more than roughly one hundred measurements, the distribution of $\tau_{ij}$ values becomes stable and its central peak becomes well resolved.  With fewer measurements, the peak remains visible, but the heavy-tailed structure is not fully developed, making robust estimation more difficult.

The third condition concerns the measurement spacing.  The decay must be slower than the time interval between spectra.  When this is true, the logarithmic differences used to compute the pairwise lifetimes smoothly change along the sequence, and their statistical variability becomes almost constant from one pair to another.  This smoothness allows the exact probability density derived by Silverman to collapse into its empirical
Cauchy-like form.

When these conditions are met, the pairwise method produces a well-defined central cluster of lifetime estimates that is ideally suited to analysis with the MFV.  The $^{97}$Ru dataset examined in this study satisfies all of these requirements: each spectrum contains high counting statistics, the time series is long, and the sampling interval is short compared with the half-life.
For this reason, the pairwise distribution is sharply peaked and the MFV provides a stable and meaningful estimate of the mean lifetime.

Although this analysis uses the same 19~K time series as that of
Goodwin \textit{et al.}~\cite{PhysRevC.80.045501}, the statistical and systematic uncertainties we quote differ from those reported in the original publication.
Systematic uncertainties depend on the analysis procedure and not solely on the underlying dataset. The MFV and profile-likelihood approaches employed here incorporate different sensitivity tests and modeling assumptions.
Therefore, the uncertainties obtained in this work should be understood as specific to this methodology rather than as revisions of the
Goodwin \textit{et al.} uncertainty budget~\cite{PhysRevC.80.045501}.

This study is restricted to a single, well-characterized the HPGe time series. The expected number of counts in a fixed counting interval follows a simple exponential decay and serves as a controlled testbed for the pairwise MFV methodology.  A natural direction for future work is to investigate how the same ideas can be extended to short-span, high-statistics trigger-rate measurements in which the observed rate is governed by a more complex model than a single exponential.

A representative example is the direct \({}^{39}\)Ar half-life measurement performed using the DEAP-3600 liquid-argon detector~\cite{adhikariDirectMeasurement39Ar2025}.  In this case, the available statistics are extremely high, and the individual rate uncertainties are correspondingly small; however, the observation window spans only a tiny fraction of the \({}^{39}\)Ar half-life.  The observable trigger rate is written as a nonlinear function of the underlying decay activity, including explicit contributions from pile-up processes, selection efficiencies, and constant background components.  Consequently, the simple closed-form pairwise lifetime estimator used in this work is not directly applicable.

However, the underlying logic of the pairwise approach remains relevant.  In a future study, pairwise or multipoint rate ratios could be constructed from the binned trigger-rate series and a toy Monte Carlo could be used to map each candidate lifetime to the expected distribution of such ratio-based proxies under the full trigger-rate model. Robust location estimators, such as the most frequent value, could then be applied to these proxy distributions as a complementary diagnostic to a global likelihood fit, particularly for identifying sensitivity to rare run-wise anomalies.

A motivation for such future studies is the recent direct
${}^{39}\mathrm{Ar}$ half-life from DEAP-3600 differs from several values previously adopted in the literature~\cite{NSR2018CH17}. The DEAP-3600 analysis reports a half-life of
\(T_{1/2}({}^{39}\mathrm{Ar}) = 302 \pm 8_{\mathrm{stat}} \pm 6_{\mathrm{syst}}\)~years, based on a high-statistics trigger-rate measurement that spans only a small fraction of the ${}^{39}\mathrm{Ar}$ half-life~\cite{adhikariDirectMeasurement39Ar2025}.
Earlier determination, based on a compilation of previous ${}^{39}\mathrm{Ar}$
half-life measurements employing various experimental techniques and subsequently re-evaluated using robust statistical methods, yield a value of
\(T_{1/2} = 268.2^{+3.1}_{-2.9}\)~years. In this re-analysis, the quoted uncertainties explicitly account for the systematic effects present in the original measurements~\cite{golovkoApplicationMostFrequent2023c}. The emergence of a statistically precise result that is in disagreement with earlier determinations highlights the importance of complementary analysis strategies that can probe DEAP-3600 model assumptions, parameter correlations, and the impact of high-statistics, short-span measurements using toy Monte Carlo data.

An additional aspect that merits investigation in future applications is the correlation structure of the trigger-rate model. In such analyses, the inferred lifetime is expected to be strongly correlated with the initial trigger-rate scale because variations in the lifetime can be partially offset by compensating changes in the overall normalization. In the original DEAP-3600 analysis~\cite{adhikariDirectMeasurement39Ar2025}, profile-likelihood methods were used to estimate parameter uncertainties, yielding a robust determination of one-dimensional confidence intervals. However, the fitted parameters' full covariance or correlation matrix was not explicitly reported. When strong parameter correlations are present, this omission can complicate the interpretation of quoted uncertainties because the apparent precision of a single parameter may depend sensitively on the chosen model parameterization and on how nuisance parameters are absorbed into the profile.

An appropriate re-parameterization of the DEAP-3600 model would help address this issue by making the dominant correlations more transparent. In particular, such a re-parameterization would allow the relative abundance of ${}^{39}\mathrm{Ar}$ in atmospheric argon to be treated as an explicit fit parameter. This quantity is of independent physical interest because the atmospheric ${}^{39}\mathrm{Ar}$/Ar ratio plays a central role in geophysical dating~\cite{jiangAr39Detection10162011a}, environmental tracer studies~\cite{luTracerApplicationsNoble2014a}, and in quantifying both cosmogenic~\cite{saldanha2019cosmogenic,zhang2022evaluation} and anthropogenic contributions to the global ${}^{39}\mathrm{Ar}$
inventory~\cite{bhattacharyaFirstExperimentalDetermination2025}. In combination with toy Monte Carlo simulations of the full trigger-rate model, this approach would allow the inferred abundance--lifetime correlation to be explored directly rather than being absorbed implicitly into the initial trigger-rate normalization.

As demonstrated in the present work for the \({}^{97}\)Ru dataset, an appropriate re-parameterization of the decay model can substantially modify the correlation structure without altering the underlying physics (refer to Section \ref{regfit}).  Applying similar re-parameterization strategies to complex trigger-rate models, followed by profile-likelihood studies of the lifetime parameter, would therefore provide a valuable and transparent cross-check of the lifetime uncertainties' robustness in future high-statistics, short-span measurements.

\section{Conclusions
}
\label{sec:conc}

A high-quality $^{97}$Ru decay dataset was re-analyzed using a method based on pairwise net-count ratios and Steiner's most frequent value statistics.  This approach removes the strong correlation between amplitude and lifetime, which affects standard regression fits, and provides a natural way to handle the heavy-tailed distribution of pairwise lifetimes.  The MFV estimate agrees closely with the regression result, but its uncertainty is defined by the behavior of the central cluster of pairwise values rather than by the amplitude model’s details, making its interpretation more transparent.

This study also clarifies the experimental conditions under which the pairwise method performs well.  When the counting statistics are sufficiently high, when the time series contains many sequential measurements, and when the sampling interval is short compared with the half-life, the distribution of pairwise lifetimes develops a sharply peaked Cauchy-like form.  Under these conditions, the MFV provides a stable and robust estimator that is resistant to the influence of extreme pairs and naturally suited to bootstrap uncertainty analysis.  The $^{97}$Ru dataset satisfies these requirements, which explains the close agreement between the MFV and regression results reported in this study.

Considering these considerations, the combination of pairwise sampling, MFV estimation, and bootstrap-based uncertainty evaluation offers a practical and reproducible analysis framework for decay measurements that meet the necessary statistical and experimental conditions.  When these conditions are satisfied, the proposed method provides a complementary perspective to traditional regression and is a useful tool for identifying heavy-tailed behavior in nuclear decay data and other time-series problems.

\authorcontributions{Conceptualization, methodology, investigation, data curation, original draft preparation, review and editing, visualization, supervision, and project administration, V.V.G. The sole author conducted all work on the manuscript and approved the final version.}

\funding{This research received no external funding.}

\dataavailability{All analysis code used in this work--including the full regression,
	profiling, correlation extraction, uncertainty budgeting, and figure
	production--is available at~\cite{Golovko2025OSF}.}

\acknowledgments{I sincerely thank Prof. Dr. John Goodwin for sharing the $^{97}$Ru half-life data. I am especially grateful for the support and assistance of Maria Filimonova. I also thank the management and staff at Canadian Nuclear Laboratories for fostering an enabling environment for this study, particularly Genevieve Hamilton and David Yuke. During the preparation of this work, I used ChatGPT (version: gpt-4o) for language checking; I subsequently reviewed and edited the content to ensure its accuracy and accept full responsibility for the final publication.}

\conflictsofinterest{The author declares no conflicts of interest.}

\newpage
\abbreviations{Abbreviations}{
	The following abbreviations are used in this manuscript:
	\\
	
	\noindent
	\begin{tabular}{@{}ll}
		MFV & most frequent value (Steiner’s robust estimator) \\
		HPGe & high-purity germanium detector for $\gamma$-ray spectroscopy \\
		EC & electron capture decay mode of $^{97}$Ru to $^{97}$Tc \\
		KL (divergence) & Kullback--Leibler divergence used in MFV information-loss minimisation \\
		GF3 & peak-fitting routine from the RADWARE suite \\
		RADWARE & software package for nuclear $\gamma$-ray spectroscopy analysis \\
		HPB & hybrid parametric bootstrap resampling framework \\
		ndf & number of degrees of freedom in $\chi^{2}/\mathrm{ndf}$ \\
		BG & background continuum under peaks (subtracted from net counts) \\
	\end{tabular}
}

\section*{Nomenclature}
\noindent
The following symbols and parameters are used in this manuscript:
\\

\small
\begin{table}[H]
	\centering
	\begin{tabularx}{\linewidth}{@{} l L @{}}
		\toprule
		\textbf{Symbol} & \textbf{Definition \hfill [units]} \\
		\midrule
		$A(t)$ & BG-subtracted net counts in a fixed counting interval at time $t$ \hfill [counts] \\
		$A_{0}$ & Activity scale at $t=0$ (initial expected net counts per interval) \hfill [counts] \\
		$N$ & Number of radioactive atoms (scale in $A(t)=\tfrac{N}{\tau}e^{-t/\tau}$) \hfill [atoms] \\
		$t$ & Time since start; $t_i,\,t_j$ denote specific measurement times \hfill [d] \\
		$y_k$ & Measured background-subtracted net peak counts for spectrum $k$ \hfill [counts] \\
		$\sigma_k$ & Statistical uncertainty of $y_k$ \hfill [counts] \\
		$\tau$ & Mean lifetime of $^{97}$Ru \hfill [d] \\
		$T_{1/2}$ & Half-life, $T_{1/2}=\tau\ln 2$ \hfill [d] \\
		$\tau_{ij}$ & Pairwise lifetime, $\displaystyle \tau_{ij}=\frac{t_j-t_i}{\ln A(t_i)-\ln A(t_j)}$ \hfill [d] \\
		$\chi^{2}$ & Chi-squared, $\displaystyle \chi^{2}=\sum_k\!\left(\frac{y_k-A(t_k)}{\sigma_k}\right)^{2}$ \hfill [--] \\
		$\Delta\chi^{2}$ & Profile-likelihood difference, $\Delta\chi^{2}(\tau)=\chi^{2}(\tau)-\chi^{2}_{\min}$ \hfill [--] \\
		$r_k$ & Normalized residual, $r_k=(y_k-A(t_k))/\sigma_k$ \hfill [--] \\
		$M$ & MFV location (the most frequent value) for $\{\tau_{ij}\}$ \hfill [d] \\
		$\varepsilon$ & MFV scale (``dihesion'') \hfill [d] \\
		$w_k$ & MFV weight, $\displaystyle w_k=\frac{\varepsilon^{2}}{\varepsilon^{2}+(\tau_k-M)^{2}}$ \hfill [--] \\
		$n_{\mathrm{eff}}$ & Effective central-count, $\displaystyle n_{\mathrm{eff}}=\sum_k w_k$ \hfill [--] \\
		$\sigma_{M}$ & MFV internal standard uncertainty, $\displaystyle \sigma_{M}=\varepsilon/\sqrt{n_{\mathrm{eff}}}$ \hfill [d] \\
		$\ln$ & Natural logarithm \hfill [--] \\
		$\text{stat},\,\text{syst}$ & Subscripts for statistical and systematic components \hfill [--] \\
		\bottomrule
	\end{tabularx}
\end{table}

\appendixtitles{yes}
\appendixstart
\appendix
\renewcommand{\theequation}{A.\arabic{equation}}
\setcounter{equation}{0}
\section{Laplace-domain consistency check
}
\label{app:laplace_plain}
\normalsize

The main results in this study are obtained in the time domain using regression, pairwise ratios, MFV summarization, and bootstrap resampling. To verify that these results do not depend on the specific time-domain fitting structure, the lifetime in the Laplace domain was also extracted using discrete sums evaluated at the measured sampling times. One Laplace-domain approach fits the projected data directly, while a second approach forms ratios in Laplace space so that the overall normalization is canceled, closely mirroring the logic of the pairwise-ratio method.

First, we define the distinct Laplace projection of the measured net counts. Let $y_k$ denote the background-subtracted net counts with standard uncertainties $\sigma_k$ at sampling times $t_k$ (in days) for the following:
$k=1,\dots,N$. For numerical stability, the time axis is shifted such that the first measurement occurs at zero,
\begin{equation}
	t_k' = t_k - \min_j t_j .
\end{equation}
This shift has no effect on the extracted lifetime and only improves exponential weight conditioning.

For a selected set of Laplace frequencies $\{s_i\}_{i=1}^{M}$ (units of
$\mathrm{d}^{-1}$), the distinct Laplace projection is defined as follows:
\begin{equation}
	\widehat{F}(s_i)=\sum_{k=1}^{N} y_k\,e^{-s_i t_k'} .
	\label{eq:disc_laplace_proj}
\end{equation}
Here, the exponential factors $e^{-s_i t_k'}$ act as positive weights, so that the projection is a weighted sum of the observed counts, with the Laplace parameter $s_i$ controlling the relative emphasis on early versus late times.
Such exponentially weighted distinct sums correspond to discrete
The Laplace transform used in stochastic and reliability analysis
\cite{ashokerallaDiscreteLaplaceTransform2025}.

The corresponding uncertainty of each projected value is propagated from the time-domain uncertainties assuming independent errors as follows:
\begin{equation}
	\sigma_{\widehat{F}}(s_i)=
	\left[\sum_{k=1}^{N}\sigma_k^2\,e^{-2 s_i t_k'}\right]^{1/2}.
	\label{eq:disc_laplace_sigma}
\end{equation}
The uncertainty is obtained using the law of uncertainty propagation for a linear combination of uncorrelated input quantities, with the exponential factors acting as weights~\cite{hughesMeasurementsTheirUncertainties2010}.

These projected data were fitted with a discrete-sum model corresponding to a single exponential decay. In this formulation, the lifetime appears only through the combination $s_i+1/\tau$,
\begin{equation}
	F_{\mathrm{model}}(s_i)=
	A_0\sum_{k=1}^{N} e^{-(s_i+1/\tau)t_k'} .
	\label{eq:disc_laplace_model_signal}
\end{equation}
To test sensitivity to any residual constant background, an optional baseline term can be added as follows:
\begin{equation}
	F_{\mathrm{model}}(s_i)=
	A_0\sum_{k=1}^{N} e^{-(s_i+1/\tau)t_k'}
	\;+\;
	B\sum_{k=1}^{N} e^{-s_i t_k'} .
	\label{eq:disc_laplace_model_signal_bg}
\end{equation}
The parameters $(A_0,\tau)$ or $(A_0,\tau,B)$ are estimated by minimizing a weighted least-squares objective in Laplace space, as follows:
\begin{equation}
	\chi^2=\sum_{i=1}^{M}
	\left(
	\frac{\widehat{F}(s_i)-F_{\mathrm{model}}(s_i)}
	{\sigma_{\widehat{F}}(s_i)}
	\right)^2 .
	\label{eq:disc_laplace_chi2}
\end{equation}

The Laplace frequencies are chosen on a logarithmic grid that spans the scale set by the decay constant. Using the initial lifetime estimate
$\tau_{\mathrm{init}}$ from the time-domain regression, we set $M=30$ and select
$s_i$ values are logarithmically spaced between $0.15/\tau_{\mathrm{init}}$ and
$6/\tau_{\mathrm{init}}$. For this dataset, this Laplace-projection fit yields a mean lifetime of $\tau = 4.0953~\mathrm{d}$. A nonparametric bootstrap with 3,000 resamples gives a central 68.27\% confidence interval of
$[4.0936,\,4.0969]~\mathrm{d}$. The corresponding half-life is given by
$T_{1/2} = 2.8386~\mathrm{d}$ with a 68.27\% interval of
$[2.8374,\,2.8398]~\mathrm{d}$. Including the baseline term
Eq.~\eqref{eq:disc_laplace_model_signal_bg} produces an indistinguishable lifetime, indicating that any residual constant background has a negligible effect.

As a complementary cross-check that is independent of the overall normalization, we also form ratios of the distinct Laplace projections at pairs of Laplace frequencies, as follows:
\begin{equation}
	R_{ij}=\frac{\widehat{F}(s_i)}{\widehat{F}(s_j)} \qquad (i<j).
\end{equation}
For a single-exponential decay, the amplitude cancels exactly, and the lifetime
$\tau$ satisfies
\begin{equation}
	\frac{\sum_{k=1}^{N} e^{-(s_i+1/\tau)t_k'}}
	{\sum_{k=1}^{N} e^{-(s_j+1/\tau)t_k'}}
	= R_{ij}.
	\label{eq:laplace_ratio_equation}
\end{equation}
This equation is numerically solved for each $(s_i,s_j)$ pair using one-dimensional root finding within a conservative bracket
$\tau\in[1,10]~\mathrm{d}$. For computational efficiency, several thousand frequency pairs are used; restricting the number of pairs does not change the result because many pairs carry redundant information.

The resulting distribution of Laplace-space lifetime estimates is summarized using the same most frequent value estimator employed in the time-domain pairwise analysis. The Laplace-ratio method yields a mean lifetime of
$\tau = 4.0952~\mathrm{d}$. The bootstrap resampling of the full time series again provides a 68.27\% confidence interval of
$[4.0935,\,4.0969]~\mathrm{d}$, corresponding to a half-life interval of
$[2.8374,\,2.8398]~\mathrm{d}$.

Uncertainties for both Laplace-domain methods are evaluated with a nonparametric bootstrap that resamples the measured time series directly. Each bootstrap replicate is created by drawing $N$ rows $(t_k,y_k,\sigma_k)$ with replacement, sorting them by time, and repeating the full Laplace-domain calculation. Using
$R=3,000$ replicates, the reported ``one-sigma'' interval is defined by the central 68.27\% of the bootstrap distribution, which corresponds to the empirical quantiles
\begin{equation}
	p_{\mathrm{lo}}=\frac{1-0.6827}{2}, \qquad
	p_{\mathrm{hi}}=1-p_{\mathrm{lo}} .
\end{equation}

Both Laplace-domain determination are fully consistent with the primary time-domain regression result of $\tau = 4.0947~\mathrm{d}$ (equivalent to $T_{1/2} = 2.8383~\mathrm{d}$) within uncertainties.
Therefore, they serve as independent robustness checks rather than as alternative estimators. The adopted lifetime and half-life values in this study are based on the time-domain analysis.

\newpage

	\PublishersNote{}
\end{document}